\documentclass[aps, pra, twocolumn, showpacs,footinbib,superscriptaddress,longbibliography]{revtex4-1}
\usepackage{graphicx, subfigure,braket}
\usepackage{amsmath,amssymb}
\usepackage{color}
\usepackage{natbib,hyperref}
\usepackage{placeins}
\usepackage{appendix}
\usepackage{booktabs,multirow,siunitx}
\usepackage[T1]{fontenc}
\begin{document}

\title{Validation of the five-electron $pO$ qubit in Si using full configuration interaction}

\author{David W. Kanaar}
\affiliation{Center for Quantum Science and Engineering, University of California, Los Angeles, 90095, California, USA}
\affiliation{Department of Mathematics, University of California, Los Angeles, 90095, California, USA}
\email{Corresponding email: dwkanaar@g.ucla.edu}
\author{John H. Caporaletti}
\affiliation{Department of Physics, University of Maryland Baltimore County, Baltimore, MD 21250, USA}
\author{J.~P. Kestner}
\affiliation{Department of Physics, University of Maryland Baltimore County, Baltimore, MD 21250, USA}
\author{Christopher R. Anderson} 
\affiliation{Center for Quantum Science and Engineering, University of California, Los Angeles, 90095, California, USA}
\affiliation{Department of Mathematics, University of California, Los Angeles, 90095, California, USA}
\author{Mark F. Gyure} 
\affiliation{Center for Quantum Science and Engineering, University of California, Los Angeles, 90095, California, USA}
\affiliation{Department of Electrical and Computer Engineering, University of California, Los Angeles, 90095, California, USA}

\begin{abstract}
The $pO$ qubit is a recently proposed charge qubit consisting of five electrons confined in a gate-defined Si quantum dot. Because its two qubit states have no dipole moment, it couples to electric fields only through the quadrupole moment, suppressing the dominant decoherence channel of conventional charge qubits while retaining all-electrical control. The original proposal was formulated in a shell-filling picture without Coulomb interactions. In this work we perform full configuration interaction (FCI) calculations to assess the impact of the Coulomb interaction on the viability of the $pO$ qubit. We find that the interacting eigenstates occupy orbital shells beyond the shell-filling picture, with the degree of excitation governed by the competition between valley splitting, orbital energy, and Coulomb energy scales. Additionally, we find that the ground states form degenerate doublets, but of two distinct kinds. Above a crossover near a valley splitting of $E_{\mathrm{VS}}\approx50\,\mu$eV the doublet is a pair of orbital angular momentum excitations with $|\langle L_z\rangle|=1$ in the lower valley, while below the crossover the doublet is a pair of single valley excitations with $\langle L_z\rangle=0$. Despite this substantial departure from the shell-filling picture, for $E_{\mathrm{VS}} > 50\,\mu$eV the essential features, vanishing dipole moments and nonzero quadrupole tensor between qubit states are preserved. We show that the vanishing dipole is in fact guaranteed by the symmetry properties of the five-electron Hamiltonian. The same energy scales that govern the orbital occupation also set the quadrupole tensor elements and the first excited state energy gap, both of which are in principle tunable through device engineering.
\end{abstract}

\maketitle

\section{Introduction}

Gate-defined semiconductor qubits are a leading platform for quantum computing, offering compatibility with existing semiconductor manufacturing processes and a small footprint that allows many qubits to be integrated on a single chip~\cite{zwerver_qubits_2022,neyens_probing_2024,dumoulin_stuyck_cmos_2026,abraham_digitally_2026}. The dominant approach within this platform encodes quantum information in the spin degree of freedom, a choice driven by the sensitivity of charge qubits to charge noise arising from their electric dipole moment~\cite{petersson_quantum_2010,kim_microwave-driven_2015}. An alternative is to keep the charge qubit states but eliminate the dipole moment such that the resulting qubit states couple through their quadrupole moment. This was first proposed in the charge quadrupole qubit~\cite{friesen_decoherence-free_2017,kornich_phonon-induced_2018,koski_strong_2020,kratochwil_charge_2021} where a single electron is shared across three quantum dots, however, this leads to low-lying leakage states. A more recently proposed alternative, the p-orbital ($pO$) qubit~\cite{caporaletti_proposed_2025,caporaletti_qubit_2026}, confines the quadrupole charge states to a single one dot, removing leakage states and giving a denser qubit architecture. This results in substantially reduced charge noise sensitivity and all-electrical qubit control, with predicted quality factors an order of magnitude beyond state-of-the-art semiconductor spin qubits. \par

The $pO$ qubit is realized by confining five electrons in a Si quantum dot. In Si, each orbital level carries a fourfold degeneracy from the combined spin and valley degrees of freedom. In a shell-filling picture, four electrons completely fill the lowest valley-degenerate $s$ orbital, leaving the fifth to occupy the $p$-orbital manifold~\cite{tarucha_shell_1996,reimann_electronic_2002,lim_spin_2011,yang_orbital_2012,leon_coherent_2020}. The two angular momentum states of this manifold, $p_+$ and $p_-$, serve as the qubit basis states. Crucially, the electric dipole matrix elements within this manifold are exactly zero. As a result, the qubit couples to the electric field only through the quadrupole tensor, via the second-order spatial derivatives of the electric potential. This eliminates the dominant decoherence channel of conventional charge qubits. Complete electrical control is achieved by modulating the shape of the quantum dot through gate voltages. The resulting qubit is predicted to have a high quality factor and fast control.\par

The original conception of the $pO$ qubit, however, neglected Coulomb interactions among the five electrons. Without interactions, the simple shell-filling picture holds exactly. Coulomb interactions can mix in configurations beyond this picture though, redistributing the electrons across higher noninteracting orbitals and potentially modifying the properties central to qubit operation. In this paper, we incorporate these interactions by performing a full configuration interaction (FCI) calculation to determine the many-body eigenstates of five electrons in a Si quantum dot, including valley splitting. The FCI approach constructs the five-electron Hamiltonian in a finite basis of Slater determinants built from single-particle orbital states and diagonalizes it exactly within this basis, capturing the orbital state configuration mixing. We restrict the calculation to the $S_z=-\frac{1}{2}$ spin sector appropriate for the $pO$ qubit.\par
Multi-electron quantum dots offer several attractive properties for hosting qubits: the core electrons can reduce charge-noise sensitivity by screening charged impurities~\cite{barnes_screening_2011, higginbotham_coherent_2014}, enhance control of the exchange interaction~\cite{yang_operation_2020}, and the larger charge distribution facilitates readout~\cite{corrigan_coherent_2021}. Coherent control has been demonstrated for electrons occupying the $p$, $d$, and $f$ shells of a Si dot~\cite{leon_coherent_2020,leon_bell-state_2021}. Describing these results quantitatively requires going beyond the single-particle picture. These multi-electron dots have been treated at the mean-field level with Hartree-Fock approaches~\cite{liang_electronic_2024,goldberg_electronic_2024} and, more accurately, with configuration interaction calculations, both without the valley degree of freedom~\cite{rontani_full_2006,barnes_screening_2011} and with it~\cite{yannouleas_valleytronic_2022,ercan_strong_2021,joecker_full_2021,dodson_how_2022,tariq_two-electron_2026}. These calculations consistently find that the Coulomb interaction mixes in configurations well beyond the naive shell-filling picture. It has been shown that the two-electron ground states of a single Si dot carry substantial weight on higher orbitals~\cite{tariq_two-electron_2026,ercan_strong_2021}. Additionally, the charge density can localize into a Wigner molecule~\cite{ercan_strong_2021,corrigan_coherent_2021}, and molecularization is strongly enhanced by anisotropy of the confinement potential~\cite{abadillo-uriel_two-body_2021}. None of these studies, however, combine the two ingredients the $pO$ qubit requires. The calculations including valleys treat two electrons in a dot~\cite{ercan_strong_2021,tariq_two-electron_2026} or donor systems~\cite{joecker_full_2021}, while those reaching five or more electrons either omit the valley degree of freedom~\cite{barnes_screening_2011} or remain at mean-field level~\cite{liang_electronic_2024}.\par

Our FCI results show that the shell-filling picture breaks down due to the Coulomb interaction as the five-electron eigenstates contain significant contributions from configurations beyond the $s$- and $p$-orbitals. Despite this, rotational symmetry about the $z$-axis and in-plane reflection symmetry of the system Hamiltonian guarantees the existence of degenerate doublets (DD) with exact parity, forbidding dipole matrix elements within.Above a valley-splitting crossover near $E_{\mathrm{VS}}\approx50\,\mu$eV, the ground state is such a doublet with $|\langle L_z\rangle|=1$, which can be controlled through a parity-conserving quadrupolar field, as required for the $pO$ qubit. Below the crossover, the ground state is instead a doublet of valley excitations with $\langle L_z\rangle=0$ that has no quadrupole coupling, making it unsuitable for the $pO$ qubit, although it could be investigated in the future in the context of valley qubits~\cite{schoenfield_coherent_2017,penthorn_two-axis_2019}.

The remainder of the paper is organized as follows. In Section~\ref{sec:Methods} we describe the FCI method used to compute the interacting five-electron eigenstates, including the construction of the single-particle orbital basis. In Section~\ref{sec:Results} we analyze the resulting eigenstates, examining how the Coulomb interaction modifies the shell-filling picture, and assess the dipole and quadrupole matrix elements between qubit states as the valley splitting is varied. In Section~\ref{sec:SF_int_spec} we present a spatial symmetry argument that identifies DDs in the spectrum with exact parity, resulting in a requirement that electric dipole matrix elements within the qubit subspace vanish exactly. Finally, Section~\ref{sec:conclusion} summarizes the main results and conclusions.

\section{Numerical Method}\label{sec:Methods}
\begin{figure}
    \centering
    \includegraphics[width=0.8\linewidth]{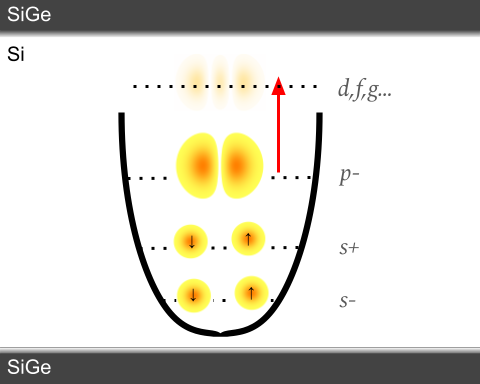}
    \caption{Schematic of a five-electron Si quantum dot with parabolic confinement potential forming the $pO$ qubit. The red arrow indicates that the Coulomb interaction pushes the electrons into higher energy non-interacting orbitals, in direct contrast to the shell-filling picture.}
    \label{fig:Schematic}
\end{figure}

We consider a idealized model of a Si quantum dot device hosting a $pO$ qubit, as schematically shown in Fig.~\ref{fig:Schematic}. The heterostructure studied here consists of a strained Si quantum well surrounded by Si$_{0.7}$Ge$_{0.3}$, though the approach is equally applicable to Si MOS devices.\par 
The five-electron $pO$ qubit states are computed using a three-step procedure following the FCI method described in Ref.~\cite{anderson_high-precision_2022}. First, the single-particle eigenstates of the system, specifically the spatial orbitals, are obtained by diagonalizing the single-particle Hamiltonian on a real-space grid. These orbitals form the many-body Slater determinant basis for the five-electron Hamiltonian. Then the Coulomb integrals for all combinations of single-particle orbitals are calculated. These Coulomb elements are used to construct the five-electron Hamiltonian, which is then diagonalized to obtain the five-electron eigenstates. \par

\subsection{Single-particle spatial orbital basis}
The single-particle states are obtained by diagonalizing the single-particle Hamiltonian evaluated on a rectangular real-space grid as in Ref.~\cite{kanaar_silicon-germanium_2026}. The grid used here has a domain of $400 \times 400 \times 10$~nm$^3$ with respective spacings $\Delta x=\Delta y=2$~nm and $\Delta z=a_0/12$ where $a_0=0.543$~nm is the Si lattice constant. The single-particle Hamiltonian takes the form
\begin{equation}
    H_{\text{sp}}(\mathbf{r})=H_{\text{kin}}(\mathbf{r})+V(\mathbf{r}),
\end{equation} 
where $H_{\text{kin}}$ is the kinetic operator, and $V$ is the potential operator. Spin-orbit coupling is neglected in this manuscript, as it is small in Si~\cite{zwanenburg_silicon_2013,burkard_semiconductor_2023,tahan_relaxation_2014,nestoklon_spin_2006}.\par
The kinetic operator is evaluated within an effective mass approximation in momentum space, where the two valleys are captured by placing minima at $k_z=\pm k_0=\pm 0.84\times 2\pi/a_0 $. Explicitly,
\begin{equation}
H_{\text{kin}}(\mathbf{k}) = \frac{\hbar^2(k_x^2 + k_y^2)}{2m_{xy}} + 
\begin{cases} 
    \dfrac{\hbar^2(k_z - k_0)^2}{2m_z} & \text{if } k_z > 0 \\[10pt]
    \dfrac{\hbar^2(k_z + k_0)^2}{2m_z} & \text{if } k_z < 0 
\end{cases}\quad,
\end{equation}
where $k_i$ and $m_{i}$ are the wavevector component and effective mass along  direction $i$, respectively. This split-operator approach is implemented efficiently using FFTs. In particular, the real-space wavefunction is Fourier transformed, $H_{\text{kin}}$ is applied, and the results are inverse Fourier transformed before applying the potential $V$ in real-space. The eigenstates of $H_{\text{sp}}$ are the effective-mass envelope of electrons times $e^{\pm ik_0z}$ for each valley, meaning all valley states are represented. These states do not include the Bloch function $u_{\pm}(r)$, which modify only the Coulomb integrals. Specifically, the Coulomb integrals contain an extra factor depending on the valley states of the single-particle basis states used. The correct Bloch function overlap between the valleys for a slowly varying potential is $\braket{u_+(r)|u_-(r)}=-0.26$~\cite{salamone_valley_2026}, leaving a factor of -0.26 for one valley mismatch and 0.26$^2$ for two mismatched valleys. These integrals are already small because they include the factors of $e^{2ik_0z}$ as a result of the valley kinetic operator. \par
The confining potential, $V(\mathbf{r})$, consists of a parabolic lateral confinement in $x$ and $y$ and a quantum well along $z$,
\begin{align}
     V(\mathbf{r})=&\frac{1}{2}m_{xy} \omega^2 (x^2+y^2)\\
       &+V_{B}\left(\frac{1}{1+e^{-4(z-z_{\mathrm{top}})/w}}+\frac{1}{1+e^{4(z-z_{\mathrm{bot}})/w}}\right)\nonumber
\end{align}
where  $\hbar \omega=1$~meV sets the lateral confinement strength\cite{ercan_strong_2021,chen_detuning_2021,mcjunkin_sige_2022,tariq_two-electron_2026}, $V_{B}=0.1506$~eV is the band offset for Si$_{0.7}$Ge$_{0.3}$~\cite{schaffler_high-mobility_1997,losert_practical_2023}, $z_{\mathrm{top}}=-z_{\mathrm{bot}}=1.5$~nm are the $z$-position of the bottom and top of the well\cite{chen_detuning_2021,acuna_coherent_2024}, and $w$ is the well interface width. The interplay between the potential and the kinetic operator sets the strength of the valley-orbit coupling, which allows the valley splitting to be tuned by choosing the well interface width~\cite{losert_practical_2023}.\par
Alloy disorder is not included in this model, though the method could easily be extended to include it. The primary reason for this omission is that obtaining meaningful results would require averaging over many random alloy disorder realizations, and a single five-electron FCI calculation is already computationally demanding due to the exponential growth of the size of the Hilbert space with electron number. \par
\begin{figure}
    \centering
    \includegraphics[width=\linewidth]{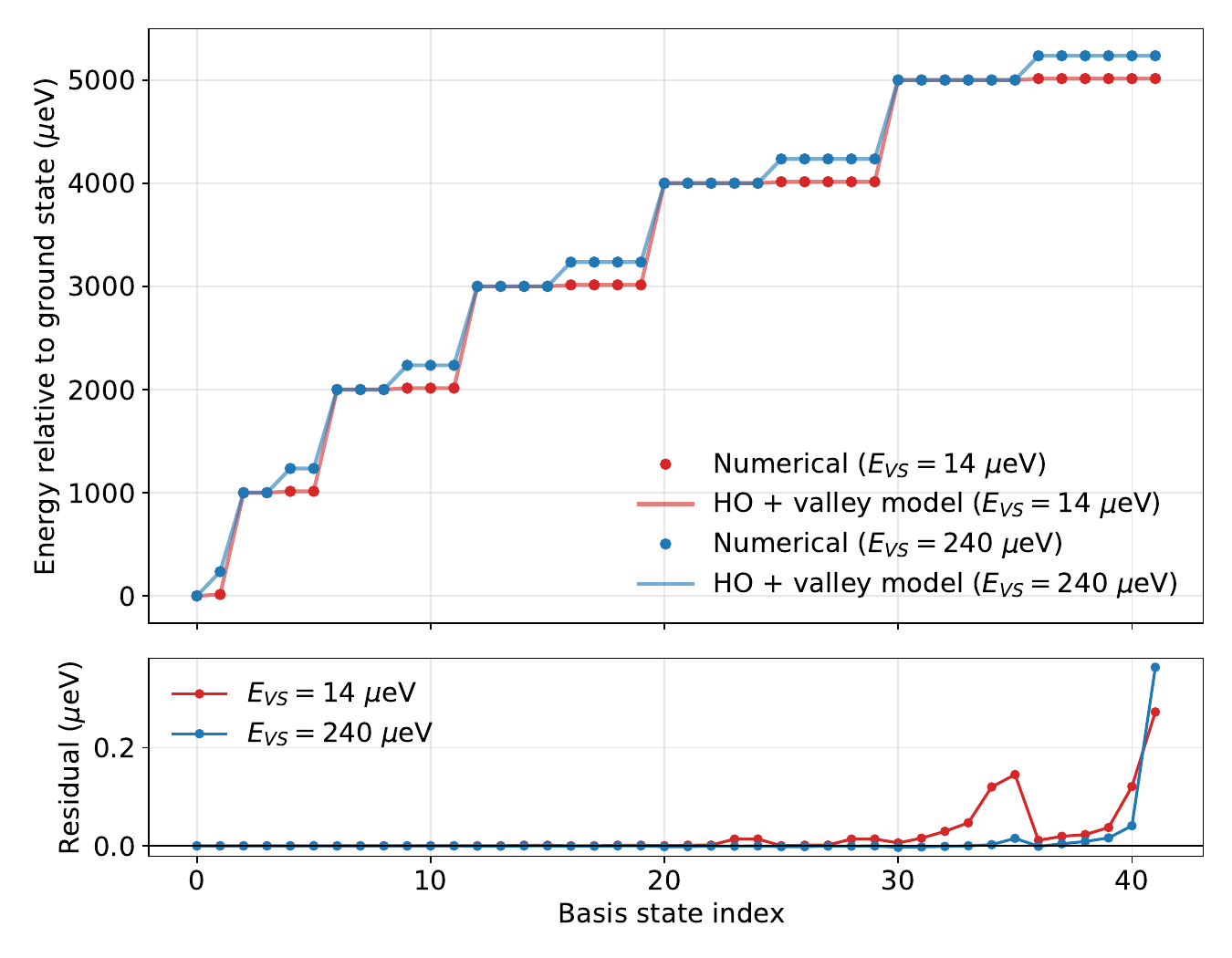}
    \caption{Top: The single-particle energies of the orbital basis used as the basis for the multi-electron slater determinants for the two different valley splitting.\\ Bottom: The difference between the analytical model for the energies of a harmonic oscillator with a valley splitting and the numerically calculated energies.}
    \label{fig:SPenergies}
\end{figure}

The single-particle Hamiltonian possesses rotational symmetry about the $z$-axis, and preserving this symmetry in the FCI results requires that complete orbital shells are retained in the single-particle basis used to construct the many-body Slater determinant basis. The numerically calculated single-particle energies align closely with those of a 2D harmonic oscillator with a valley splitting, as shown in Fig.~\ref{fig:SPenergies}, confirming the shell structure of the single-particle basis. The many-body Hilbert space scales as $N^5$ with the number of spatial orbitals $N$, placing practical constraints on the basis size. Memory limitations restrict our calculations to including up to the $h$ shell of orbitals, giving a total of 42 spatial basis states. 

\subsection{Multi-particle Hamiltonian and FCI method}
The five-electron Hamiltonian, $H_{\mathrm{5e}}$, is constructed as the sum of single-particle Hamiltonians and the pairwise Coulomb interactions,
\begin{equation}\label{eq:H5e}
    H_{\mathrm{5e}}(\mathbf{r}_1,..,\mathbf{r}_5)=\sum_i^5H_{\text{sp}}(\mathbf{r}_i)+\sum_{i>j}^5 \frac{e^2}{4\pi \epsilon|\mathbf{r}_i-\mathbf{r}_j|},
\end{equation}
where $e$ is the electron charge and $\epsilon=\epsilon_r\epsilon_0$ is the permittivity, taken throughout the computational region to be $\epsilon_r=11.7$ for Si. The difference in dielectric constant between Si and Si$_{0.7}$Ge$_{0.3}$ is neglected. This is justified because they only differ by approximately 10\%, and the large band offset between the Si well and the Si$_{0.7}$Ge$_{0.3}$ buffer causes the wavefunctions to exponentially decay in the buffer region.

The five-electron Hamiltonian, $H_{\mathrm{5e}}$, is expressed in the basis of Slater determinants constructed from the single-particle spatial orbitals combined with the spin $\uparrow$ and $\downarrow$ states. Diagonalizing within this basis following the method of Ref.~\cite{anderson_rayleighchebyshev_2010} yields the many-body eigenstates.\par Since the $pO$ qubit is ultimately a charge qubit, the charge and transition charge densities are of particular interest. The charge transition density between eigenstate $i$ and $j$ is given by
\begin{equation}\label{eq:ChargeDensity}
    \begin{split}
     \rho_{ij}(\mathbf{r})=& \, 5\int\Psi_i(\mathbf{r},\mathbf{r}_2,..,\mathbf{r}_5)\Psi_j^*(\mathbf{r},\mathbf{r}_2,..,\mathbf{r}_5)d\mathbf{r}_2..d\mathbf{r}_5,
\end{split}
\end{equation}
where the factor of 5 accounts for the indistinguishability of the electrons. The charge density of eigenstate $i$ is denoted by $\rho_{ii}\equiv \rho_{i}$. The transition charge density  $\rho_{01}$ between the charge densities of the ground states determines the moments that couple qubit states.
Throughout this work we restrict the calculations to the spin sector $S_z=-\frac{1}{2}$, as no spin-flipping processes are expected on the timescales of the $pO$ qubit controls, so all spin integrals are performed analytically and omitted from the expressions.

\subsection{Shell-filling reference and occupation measures}\label{subsec:ShellFillandMeasures}

The non-interacting shell-filling picture introduced above serves as a reference for the FCI results in Sec.~\ref{sec:Results}. The six reference states compared with the six lowest FCI eigenstates place the fifth electron in the lower-valley  $p$ orbitals, the upper-valley  $p$ orbitals, and the lower-valley $d$ orbitals while the four other electrons always fill the $s$ shell. Where shell-filling charge densities and moments are needed, they are constructed from the $E_{\mathrm{VS}}=240\,\mu$eV single-particle orbitals. 

The single-particle Hamiltonian eigenstates (Fig.~\ref{fig:SPenergies}) closely follow those of a 2D harmonic oscillator with a low energy valley degree of freedom. As a result, each orbital can be assigned a shell index $\alpha\in\{s,p,d, ...\}$ with quantum number $N_\alpha=0,1,2,...$, containing $N_\alpha+1$ degenerate orbitals per valley $\nu\in\{-,+\}$. These have the energies  $\approx\hbar\omega(N_\alpha+1)$, shifted by $E_{\mathrm{VS}}$ for $\nu=+$. The occupation of a shell $\alpha$ in valley $\nu$ is denoted by $n_\alpha^\nu$, the expectation value of the number operator summed of the orbitals of that shell. To quantify the deviation of the interacting eigenstates from the shell-filling picture, we define the of the orbital and valley excitation
\begin{align}
    &\langle N_{\mathrm{orb}} \rangle = \sum_{\alpha,\nu} n_\alpha^\nu N_{\alpha}\quad ,\\
    &\langle  \mathrm{valley} \rangle = \sum_{\alpha} n_\alpha^+ ,
\end{align}
which equal $1$ and $2$, respectivly, for the shell-filling ground states.

Additionally, the orbital angular momentum expectation value $\langle L_z\rangle$ is evaluated by contracting the one-body reduced density matrix of each eigenstates with the single-particle matrix elements of $L_z=-i\hbar (x \partial_y-y\partial_x)$. It is reported in units of $\hbar$.

\section{Results}\label{sec:Results}
\begin{figure}[h]
    \centering
    \includegraphics[width=\linewidth]{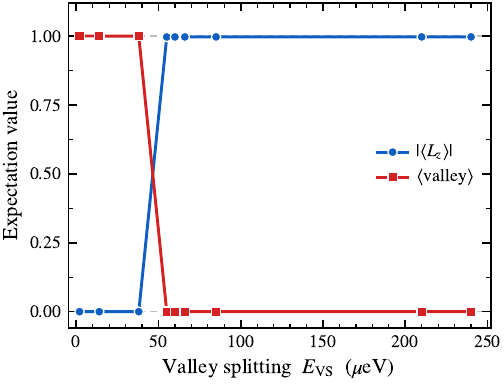}
    \caption{Valley excitation expectation, $\langle \mathrm{valley} \rangle$, and orbital angular momentum, $|\langle L_z\rangle|$, of the FCI ground state manifold as a function of valley splitting.}
    \label{fig:LZandValleyregime}
\end{figure}

In this section we present and analyze the six lowest energy eigenstates of the five-electron Hamiltonian obtained from the FCI calculations described in Section~\ref{sec:Methods}. The eigenstates are computed for nine valley splittings between $E_{\mathrm{VS}}=2\,\mu$eV and $240\,\mu$eV, obtained by varying the interface widths $w$ listed in Table~\ref{tab:vs_widths}. In every case the ground state manifold is a degenerate doublet (DD), but it is not always the same doublet. Two distinct regimes were found with a crossover near $E_{\mathrm{VS}}\approx50\,\mu$eV and distinguished by the orbital angular momentum, $\langle L_z\rangle$, and valley excitation,  $\langle\mathrm{valley}\rangle$ as shown in Fig.~\ref{fig:LZandValleyregime}. In the high-valley-splitting regime, $E_{\mathrm{VS}}\gtrsim 50\,\mu$eV, the ground doublet carries $|\langle L_z\rangle|=1$ and $\langle\mathrm{valley}\rangle=0$.  In the low-valley-splitting regime, $E_{\mathrm{VS}} \lesssim50\,\mu$eV, the ground doublet carries $\langle L_z\rangle=0$ and $\langle\mathrm{valley}\rangle=1$, making it a pair of valley excitations with no net orbital angular momentum. In the remainder of this section the two regimes are represented by $E_{\mathrm{VS}}=240\,\mu$eV and $14\,\mu$eV respectively, while the full results for each run are given in Appendix~\ref{app:allValleyCases}.

\begin{table}[]
    \centering
    \begin{tabular}{c|c}
        $E_{\mathrm{VS}}$ ($\mu$eV) & $w$ (nm) \\
        \hline
         2   &  0.679  \\
         14  &  0.543 \\
         38  &  0.448\\
         55  &  0.421\\
         60  &  0.407  \\
         66  &  0.394\\
         85  &  0.367 \\
         210 &  0.299\\
         240 &  0.272 \\
    \end{tabular}
    \caption{Valley splitting values $E_{\mathrm{VS}}$ used in the FCI calculations and their corresponding interface widths $w$ rounded to three significant figures.}
    \label{tab:vs_widths}
\end{table}
We first examine the dependence of the eigenstate energies on valley splitting, before turning to the charge densities and orbital occupations defined in Sec.~\ref{subsec:ShellFillandMeasures}, and finally, we examine the dipole moments and quadrupole tensors, which determine the coupling of the qubits states to electric fields and set the scale for both gate speed and charge noise sensitivity. Taken together, the results show that while the Coulomb interactions push the electrons into much higher orbitals than the shell-filling model predicts, in the high-valley-splitting regime, the properties essential for $pO$ qubit operation are preserved.

\subsection{FCI eigenstate energies}
\begin{figure}[h]
    \centering
    \includegraphics[width=\linewidth]{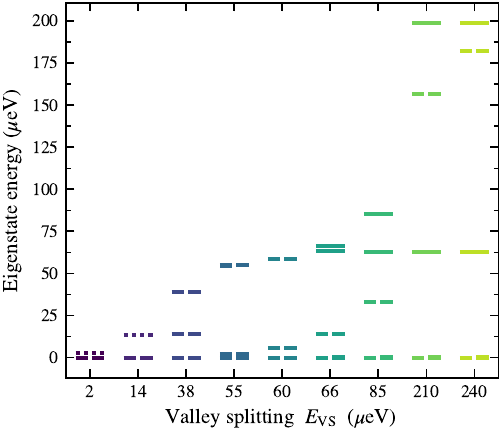}
    \caption{Energies of the six lowest energy eigenstates of Eq.~\eqref{eq:H5e} for all valley splittings. Near degenerate levels are indicated by multiple dashes at an energy level line.}
    \label{fig:FCIenergies}
\end{figure}

Fig.~\ref{fig:FCIenergies} shows the energies of the lowest six eigenstates for the different valley splittings. For each case, the ground state is a DD manifold. As expected, the energy gap to the first nondegenerate state increases as the valley splitting increases in the high-valley-splitting regime. Of particular note is that the gap to the first excited state is relatively small even for the highest valley splitting, $E_{\mathrm{VS}}=240\mu$eV, with an energy gap of $\approx63\mu$eV. This is about 7 times the thermal energy of $k_B T\approx8.6\mu$eV at an electron temperature of $T=$100 mK~\cite{petit_spin_2018} where $k_b$ is Boltzmann's constant. While this energy gap is modest, Fig.~\ref{fig:FCIenergies} makes it clear that this gap is set by the valley splitting in competition with the orbital energy and Coulomb energy. Both the valley splitting and orbital energy are tunable in principle for a device. This likely means that the next-excited-state energy gap can be pushed towards the standard threshold of $100\mu$eV\cite{esposti_confinement_2026} through device engineering. 

\subsection{Charge densities}
\begin{figure*}[t!]
    \centering
    \includegraphics[width=0.95\linewidth]{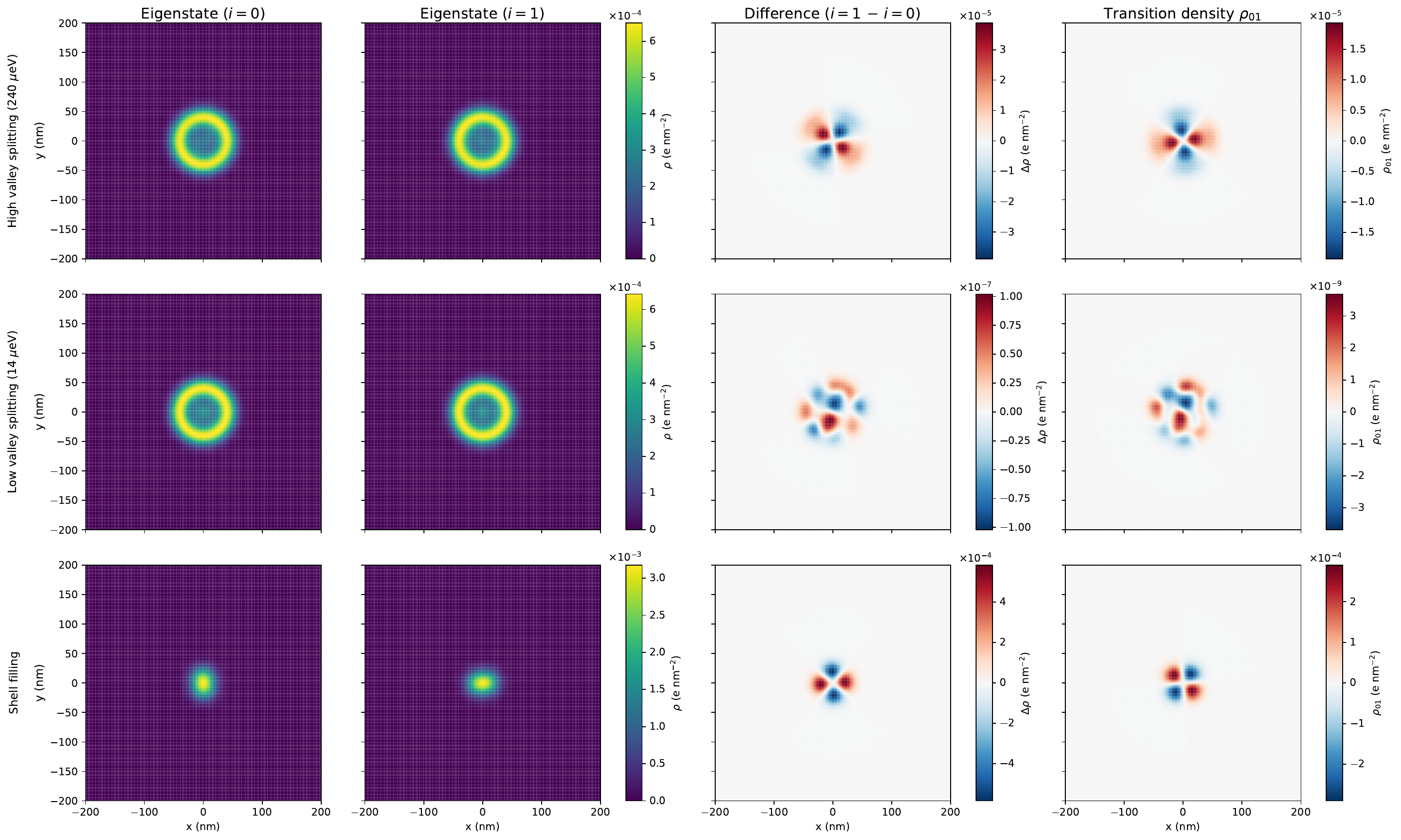}
    \caption{The charge densities, integrated over $z$, of the many-body ground states, their difference, and the transition between them (columns), for three different cases, (rows): high-valley-splitting regime ($E_{\mathrm{VS}}$=240\, $\mu$eV), low-valley-splitting regime ($E_{\mathrm{VS}}$=14\, $\mu$eV), and a shell-filling model using the orbital basis of $E_{\mathrm{VS}}$=240 $\mu$eV.}
    \label{fig:2DChargedensity}
\end{figure*}

The many-body charge densities of the DD ground state manifold are computed using Eq.~\eqref{eq:ChargeDensity} for the interacting cases and for a shell-filling model with $E_{\mathrm{VS}}=240\,\mu$eV. Fig.~\ref{fig:2DChargedensity} shows the charge densities integrated over $z$ for the two regimes and the shell-filling case. The shell-filling and high-valley-splitting regime charge densities, shown in the bottom and top row of Fig.~\ref{fig:2DChargedensity}, capture the original $pO$ qubit picture clearly: the density differences and transitions are between $p_x$- and $p_y$-like orbitals with minimal spatial overlap, resulting in a large charge difference and thus a large quadrupole moment. \par In contrast, in the low-valley-splitting regime both the ground-state charge density difference and the transition charge density are negligible. In fact, the color scale in the middle row of Fig.\ref{fig:2DChargedensity} has a range five orders of magnitude smaller than in the other cases and even then shows no meaningful structure. This means the low-valley-splitting regime is not suitable for the $pO$ qubit as there is no quadrupole coupling between the states as quantified in Sec.~\ref{subsec:moments}. Additionally, all interacting charge densities are considerably more spatially extended, with the degree of spreading increasing with valley splitting. This trend is physically reasonable because the valley splitting increases the energy penalty for occupying excited-valley orbitals, making it more favorable for Coulomb repulsion to push electrons to many different higher orbitals with lower valleys that are more spread out. Consequently, the amplitude of the FCI charge density difference is reduced compared to the shell-filling model and the characteristic $p$-orbital structure is also spread out. This foreshadows the decrease in quadrupole tensor elements relative to the shell-filling picture quantified in Section.~\ref{subsec:moments}.

\subsection{Orbital occupation}
\begin{figure}[h]
    \centering
    \includegraphics[width=\linewidth]{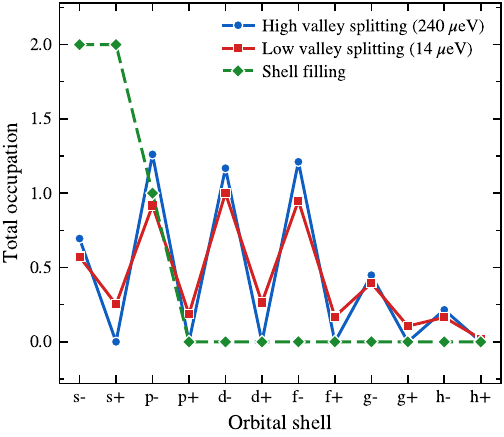}
    \caption{Orbital shell occupation of the multi-electron ground state for each of the two FCI regimes as well as in the shell-filling picture.}
    \label{fig:ShellOccupation}
\end{figure}

Figure~\ref{fig:ShellOccupation}. shows the shell occupations $n_{\alpha}^\nu$ of the ground doublet for the two valley-splitting regimes and for the shell-filling picture.

Compared to the shell-filling picture, the FCI ground states for both regimes spread across a large number of orbital shells. Correspondingly, the ground state doublets have $\langle N_{\mathrm{orb}} \rangle\approx10$ in both regimes, far exceeding the shell-filling picture of $\langle N_{\mathrm{orb}} \rangle=1$, with a slightly larger value for the high-valley-splitting regime. In contrast to the orbital excitation, $\langle \mathrm{valley} \rangle$ remains close to an integer and distinguishes the cases clearly. For the ground states it is $2$ in the shell-filling picture, $1$ in the low-valley-splitting regime, and $0$ in the high-valley-splitting regime. The total valley expectation values remain close to integer values across all eigenstates because the intervalley Coulomb coupling is very weak. After the crossover valley splitting, the energy penalty for occupying the upper valley exceeds the Coulomb cost of confining all electrons to the lower valley and the electrons spread to slightly higher shells instead. Below the crossover valley splitting, in the low-valley-splitting regime, the single valley excitation is spread across a much broader range of orbital shells compared to the shell-filling picture. The orbital and valley expectation values for all states and valley splitting cases are plotted in Appendix~\ref{app:allValleyCases}.\par

 Taken together, the valley, orbital, and angular momentum (shown in Fig.~\ref{fig:LZandValleyregime}) trends characterize the degenerate doublet that forms the $pO$ qubit states. The composition of these states is governed by the interplay between the orbital, valley, and Coulomb energy scales, and deviates significantly from the noninteracting shell-filling picture. In the low-valley-splitting regime the energy competition results in the ground doublet with $\langle L_z\rangle=0$. Since both states then have the same angular momentum, there is no in-plane charge density difference between them and therefore no quadrupole coupling, making that regime unsuitable for operating the $pO$ qubit. However, this is not a practical restriction, since a valley splitting of at least $\sim 100\,\mu$eV is already required for spin qubit devices to survive initialization, readout, and thermal excitations\cite{culcer_realizing_2009}. 

The key properties of the $pO$ qubit are not the orbital occupations \textit{per se}, but the properties of the charge density that govern the coupling to the electric field, namely a vanishing dipole moment and a sufficiently large quadrupole tensor, which we examine next.

\subsection{Magnitudes of the dipole moments and quadrupole tensor} \label{subsec:moments}
The transition dipole moment between eigenstates $i$ and $j$ is computed as 
\begin{equation}\label{eq:transitiondipole}
    \mathbf{d}_{ij}=\int \rho_{ij}(\mathbf{r})\,\mathbf{r}\,d \mathbf{r},
\end{equation}
and the permanent dipole moment of state $i$ is denoted as $\mathbf{d}_{i}=\mathbf{d}_{ii}$. In this section, the states of interest are the $pO$ qubit states, the lowest two five-electron eigenstates calculated using FCI, labeled $i=0$ and $i=1$.
As in the simple shell-filling picture, the magnitudes of the dipole moment difference, $\Delta \mathbf{d} = \mathbf{d}_{1} - \mathbf{d}_{0}$, and transition dipole moment, $\mathbf{d}_{01}$, are found to be zero within the numerical precision of our calculation (more details are listed in Appendix~\ref{app:Moments}). This is not a coincidence. As shown in Section~\ref{sec:SF_int_spec}, the symmetries of the Hamiltonian require the dipole moments to vanish exactly even in the strongly interacting case.\par

For the control of the $pO$ qubit, it is essential that the quadrupole tensor differs between the qubit states and that the transition quadrupole between them must be nonzero, as these drive the logical $Z$ and $X$ gates. The quadrupole tensor, $Q_{ij}$, associated with the (transition) charge densities of Eq.~\eqref{eq:ChargeDensity} is
\begin{equation}
    Q_{ij}=\int \rho_{ij}(\mathbf{r})\,\left(3\,\mathbf{r} \otimes\mathbf{r} -r^2 \mathbf{I} \right)\,d \mathbf{r},
\end{equation}
where $\otimes$ is the outer product and $\mathbf{I}$ is the $3 \times 3$ identity operator.  As with the dipole moments, $Q_{ii}$ is the quadrupole tensor of state $i$ and $Q_{01}$ is the transition quadrupole between the qubit states. The Frobenius norms of the quadrupole difference $\Delta Q=Q_{11}-Q_{00}$ and transition quadrupole $Q_{01}$ for all cases, including the shell-filling picture, are listed in Table~\ref{tab:NormQuadrupole} while the full tensors are given in Appendix~\ref{app:Moments}. Both norms are nonzero and nearly valley-splitting independent in the high-valley-splitting regime, but vanish in the low-valley-splitting regime. As noted in the previous subsection, this is because the two ground states at low valley splitting become a $\langle L_z\rangle=0$ valley doublet with almost identical charge densities instead of an angular momentum doublet with $|\langle L_z\rangle|=1$ with a permitted quadrupole coupling.

\begin{table}[h]
    \centering 
    \caption{Frobenius norms of the quadrupole tensors of the difference of the two logical states, and for the transition density between them (in units of $\,e \cdot \text{nm}^2$).}
    \begin{tabular}{|l|c|c|}
        \hline
        Case  & $\|\Delta Q\|$ & $\|Q_{01}\|$ \\ \hline
        $E_\text{VS} = 2~\mu\text{eV}$   &   0.076 &   0.032 \\
        $E_\text{VS} = 14~\mu\text{eV}$  &   0.063 &   0.001 \\
        $E_\text{VS} = 38~\mu\text{eV}$  &   0.086 &   0.044 \\ \hline
        $E_\text{VS} = 50~\mu\text{eV}$  & 360.636 & 180.324 \\
        $E_\text{VS} = 60~\mu\text{eV}$  & 360.743 & 180.391 \\
        $E_\text{VS} = 66~\mu\text{eV}$  & 360.646 & 180.441 \\
        $E_\text{VS} = 85~\mu\text{eV}$  & 360.823 & 180.339 \\
        $E_\text{VS} = 210~\mu\text{eV}$ & 360.901 & 180.341 \\
        $E_\text{VS} = 240~\mu\text{eV}$ & 360.720 & 180.497 \\ \hline
        Shell                            & 1701.517 & 850.758 \\ \hline
    \end{tabular}\label{tab:NormQuadrupole}
\end{table}

The quadrupole tensor difference norms, listed in Table~\ref{tab:NormQuadrupole}, for the interacting cases decrease as compared to the shell-filling picture. Although the norm of the quadrupole tensor difference between basis states generally increases at larger basis index, see Appendix~\ref{app:BasisMoments}, the interacting eigenstates are a superposition of many orbitals. Because the quadrupole tensors of the contributing orbitals are generally oriented in different directions the resulting quadrupole tensor difference for the five-electron eigenstate decreases. This is also visible in the ground state differences shown in Fig.~\ref{fig:2DChargedensity} are more spread out in the high-valley-splitting regime leading to a lower quadrupole moment. 

Even though $\|\Delta Q\|$ is reduced by a factor of about five compared to the shell-filling prediction, we find that the remaining values are sufficient to achieve fast driving. To estimate the Rabi frequency, we project the quantum dot perturbation $\sum_i\frac{1}{2}m_{xy}(\omega_x^2x_i^2-\omega_y^2y_i^2)$ into the ground state manifold and calculate the energy splitting to obtain
\begin{equation}
    f_R=\frac{m_{xy}}{2h}(\omega_x^2\hspace{-0.2em}-\hspace{-0.2em}\omega_y^2)\sqrt{\|\Delta Q/2\|^2\hspace{-0.2em}+\hspace{-0.2em}\| Q_{01}\|^2}\hspace{-0.2em}\approx\hspace{-0.2em}10\hspace{-0.2em}~\text{GHz},
\end{equation}
where $\hbar\omega_x=1~\text{meV}$ and $\hbar \omega_y=1.1~\text{meV}$, such that $\hbar (\omega_x-\omega_y)=100~\mu\text{eV}$~\cite{caporaletti_proposed_2025}, and we use the FCI quadrupole tensor norms in Table~\ref{tab:NormQuadrupole} when $|\langle L_z\rangle|=1$. Since the dipole moments vanish exactly by symmetry, the quadrupole tensor elements also govern the coupling to electric field fluctuations. This means that the quality factor as a result of these fluctuations is not changed by having a smaller $\|\Delta Q\|$ as driving speed is also proportional to this. The nonzero values obtained here therefore indicate that coherent control of the $pO$ qubit remains feasible despite significant departure from the shell-filling picture.

\section{Symmetry Arguments for the Existence of Degenerate, Dipole-Free Doublets} \label{sec:SF_int_spec}

The dipole moment and splitting of the $pO$ qubit computed in Section~\ref{sec:Results} are zero within the numerical precision of our simulation. This is not a coincidence, but rather a direct consequence of the system's spatial symmetry.

The degeneracy originates from the rotational symmetry of the system Hamiltonian about the $z$-axis. That is, the total $z$ orbital angular momentum operator $L_z$ commutes with the system Hamiltonian from Eq.~\eqref{eq:H5e},
\begin{equation}\label{eq: rot_sym}
    [H_{5e},L_z]=0.
\end{equation}
This guarantees that $H_{5e}$ and $L_z$ have a common set of eigenstates, denoted $\ket{m}$, where $L_z\ket{m}=\hbar m\ket{m}$ and $H_{5e}=E_m\ket{m}$. We now prove that eigenstates with the same magnitude of $z$ angular momentum $|m|$ are degenerate using the in-plane reflection operator $\Pi_x$.

The $\Pi_x$ operator reflects each particle's coordinates through the $yz$-plane, i.e., $x_i\rightarrow-x_i$, commutes with the system Hamiltonian $[H_{5e},\Pi_x]=0$, and anti-commutes with the total orbital $z$ angular momentum $\{L_z,\Pi_x\}=0$. From these properties, it follows that
\begin{equation}\label{eq:DegDoub}
\begin{aligned}
    &H_{5e}\ket{-m}=H_{5e} \Pi_x \ket{m}=\Pi_x H_{5e}\ket{m}\\
    &=\Pi_x E_m\ket{m}=E_m\ket{-m}.
\end{aligned}
\end{equation}
Therefore, every pair $\ket{\pm m}$ is degenerate (with the obvious exception of $m=0$, since that is only one state). The $pO$ qubit is therefore expected to be degenerate because it is composed of the states $\ket{\pm1}$.

The lack of dipole moment originates from the system's parity symmetry, where the parity operator $\Pi$ transforms each particle's coordinates by $x_i\rightarrow -x_i$, $y_i\rightarrow-y_i$, $z_i\rightarrow -z_i$ and can be decomposed into a reflection through the $xy$ plane followed by a $\pi$ rotation about the $z$-axis
\begin{equation}\label{eq:parity}
    \Pi=R_z(\pi)\Pi_z,
\end{equation}
where $R_z(\pi)= e^{i \pi L_z/\hbar}$ and $\Pi_z$ takes $z_i\rightarrow-z_i$. Since
\begin{equation}
    [H_{5e},L_z]=[H_{5e},\Pi]=[L_z,\Pi]=0,
\end{equation}
we can add one more quantum number to the system eigenstates such that $\ket{m}\rightarrow\ket{m,z_p}$, where
\begin{equation}\label{eq:Rz,Piz}
\begin{aligned}
    &\Pi_z\ket{m,z_p}=z_p\ket{m,z_p},\\
    &R_z(\pi)\ket{m,z_p}=e^{i\pi m}\ket{m,z_p},
\end{aligned}
\end{equation}
and $z_p=\pm1$ is the eigenvalue of $\Pi_z$.

Using Eq.~\eqref{eq:parity} and Eq.~\eqref{eq:Rz,Piz}, we now project the dipole operator into the basis states $\ket{m,z_p}$,
\begin{multline}\label{eq:dip_0}\sum_k\braket{m^\prime,z_p^\prime|\mathbf{r}_k|m,z_p}=\sum_k\braket{m^\prime,z_p^\prime|\Pi^\dagger\Pi\mathbf{r}_k\Pi^\dagger\Pi|m,z_p}\\
    =\sum_k-z_p^\prime z_p e^{i\pi(m^\prime+m)}\braket{m^\prime,z_p^\prime|\mathbf{r}_k|m,z_p},
\end{multline}
where $r_k$ is the position operator for the $k$-th electron and transforms under parity as $\Pi\mathbf{r}_k\Pi^\dagger=-\mathbf{r}_k$.
Restricting the matrix element in Eq.~\eqref{eq:dip_0} to a DD derived in Eq.~\eqref{eq:DegDoub}, term $k$ must satisfy
\begin{equation}\label{eq:dip_0_v2}
    \braket{m^\prime,z_p^\prime|\mathbf{r}_k|m,z_p}=-e^{i\pi(m^\prime+m)}\braket{m^\prime,z_p^\prime|\mathbf{r}_k|m,z_p}
\end{equation}
whose solution is zero for $m^\prime=-m$, corresponding to the doublets transition dipole, and $m^\prime=m$, the permanent dipole, and we assumed that $z_p^\prime=z_p$ because states with different $z$ parity are generally non-degenerate for nonzero valley splitting. Therefore, each DD has zero dipole moment.

Our symmetry analysis holds for any number of electrons because the Coulomb interaction preserves rotational symmetry. Therefore, the argument generally states that multi-electron, DD, zero-dipole manifolds spanned by $\ket{\pm m}$ exist in the system's spectrum. However, it does not determine if any of these manifolds are the ground state. That depends on electron number, confinement energy, and valley splitting. We choose to use five electrons in the interacting FCI simulation simply because in the shell-filling case, and the limit of small valley splitting, five electrons are needed to guarantee the ground state is a $pO$ qubit manifold. Interestingly, Sec.~\ref{sec:Results} finds that when interactions are considered, the $pO$ qubit manifold is the ground state for a five-electron dot for \emph{large} enough valley splitting rather than small valley splitting. We attribute this deviation from the shell-filling picture to the strong e-e interactions. 

As an aside, note that the ground state DD manifold when $E_{\mathrm{VS}}\lesssim50~\mu\text{eV}$ also has zero dipole moment, but it is not subject to our symmetry argument above, so there must be some additional symmetry principle at play there.

Unlike the dipole moment, the quadrupole tensor is even under parity inversion because each element scales as $r_i r_j$ so $\Pi r_i r_j\Pi^\dagger=(-r_i )(-r_j)=r_i r_j$ and is therefore not required to vanish. However, the quadrupole tensor still follows angular selection rules, i.e., it couples states which differ by $\pm2$ units of $z$ orbital angular momentum. Therefore, only the $pO$ states spanned by $\ket{\pm1}$ can be driven through their quadrupole moment.

\section{Discussion and conclusion}\label{sec:conclusion}

The five-electron FCI results demonstrate that the Coulomb interaction substantially modifies the $pO$ qubit eigenstates relative to the shell-filling picture, pushing the electrons to occupy higher orbital shells. A clear picture emerges of two distinct valley-splitting-dependent regimes distinguished by the character of the degenerate doublet (DD) ground states. Below a valley splitting crossover of about $E_{\mathrm{VS}}\approx 50\,\mu$eV, the low-valley-splitting regime, the doublet is a pair of valley excitations, with $\langle \mathrm{valley} \rangle = 1$ and $\langle L_z \rangle = 0$. Above the crossover, the high-valley-splitting regime, the doublet is a pair of orbital angular momentum excitations, with $|\langle L_z \rangle| = 1$ and $\langle \mathrm{valley} \rangle = 0$. The competition between the valley splitting, orbital energy, and Coulomb energy scales sets the transition point between these regimes and the degree of spreading into higher shells. Despite this departure, the dipole moments remain zero to within numerical precision meaning the $pO$ qubit maintains low charge-noise sensitivity. Additionally, the quadrupole tensor elements are reduced relative to the shell-filling picture, and in the low-valley-splitting regime they vanish. In the high-valley-splitting regime these elements are large enough for high fidelity qubit control. The $pO$ qubit must therefore be operated above the valley-splitting crossover, which for $\hbar \omega = 1$meV we find near $50~\mu$eV. Valley splittings of $\sim 100~\mu$eV are regularly achieved in Si/SiGe heterostructures, placing realistic devices comfortably within the operating regime~\cite{borselli_measurement_2011,degli_esposti_low_2024,esposti_confinement_2026,abraham_digitally_2026,marcks_valley_2025,huang_high-fidelity_2024}.\par

The degeneracy and vanishing dipole are a direct consequence of the spatial symmetries of the five-electron Hamiltonian. Rotational symmetry around the $z$ axis together with in-plane reflection symmetry guarantee that $L_z$ eigenstates with $\pm m$ form degenerate doublets. Additionally, parity symmetry forces the permanent and transition dipole moments within the qubit doublet to vanish exactly, independently of electron number or interaction strength. The role of having five electrons is to place a DD with $|\langle L_z\rangle|=1$ at the bottom of the spectrum, which is what quadrupolar control requires~\cite{leon_bell-state_2021}. However, this holds only in the high-valley-splitting regime where an excited valley DD is not the ground state. The vanishing dipole rests on parity alone and therefore survives perturbations such as dot ellipticity that merely lift the rotational symmetry, whereas any parity-breaking perturbation, for example interface disorder, will generically reintroduce a dipole moment.\par

The FCI details further inform how device parameters shape the qubit properties. Higher valley splitting suppresses upper-valley occupation and spreads electrons across higher lower-valley orbitals, directly affecting the quadrupole tensor difference and the first excited state energy gap. Both the excited state gap and the quadrupole moment are in principle tunable through device engineering of the valley splitting and lateral confinement, suggesting a path toward optimizing gate speed and noise properties simultaneously within an experimentally realistic parameter range.

\section*{Acknowledgments}
Research was sponsored by the Army Research Office and was accomplished under Grant Number W911NF-23-1-0115.  The views and conclusions contained in this document are those of the authors and should not be interpreted as representing the official policies, either expressed or implied, of the Army Research Office or the U.S. Government. The U.S. Government is authorized to reproduce and distribute reprints for Government purposes notwithstanding any copyright notation herein. This work used computational and storage services associated with the Hoffman2 Cluster which is operated by the UCLA Office of Advanced Research Computing’s Research Technology Group.
\appendix

\section{Results for all valley splitting cases}\label{app:allValleyCases}
\begin{figure}[h]
    \centering
    \includegraphics[width=\linewidth]{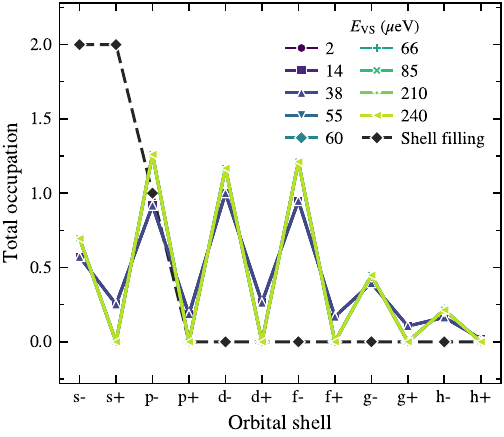}
    \caption{Orbital shell occupation of the multi-electron ground states for all FCI calculations and for the shell-filling picture.}
    \label{fig:ShellOccupationAll}
\end{figure}

\begin{figure}[h]
    \centering
    \includegraphics[width=\linewidth]{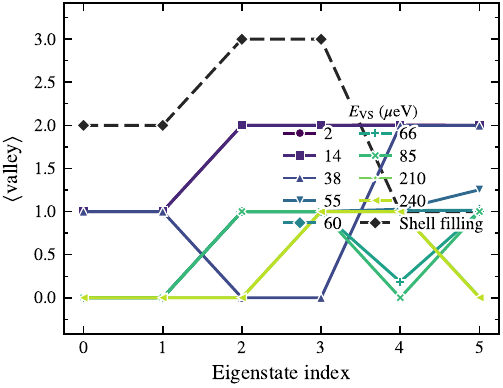}
    \caption{Valley expectation of the lowest six multi-electron states for all FCI calculations and for the shell-filling picture.}
    \label{fig:ValleyExpectationsAll}
\end{figure}

\begin{figure}[h]
    \centering
    \includegraphics[width=\linewidth]{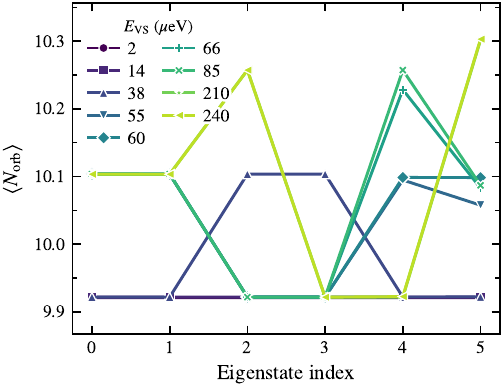}
    \caption{Orbital expectation of the lowest six multi-electron states for each of all FCI results as well as for the shell-filling picture.}
    \label{fig:OrbitalExpectationAll}
\end{figure}

In this section the occupation, valley expectation, and orbital expectation results are shown including all valley splitting cases in Figs.~\ref{fig:ShellOccupationAll}, \ref{fig:ValleyExpectationsAll}, and \ref{fig:OrbitalExpectationAll}. 


\section{Full dipole moments and quadrupole tensors} \label{app:Moments}
In this section of the appendix we report the full quadrupole and dipole moments computed from the FCI charge densities as well as the charge density according to the shell-filling model using the 240~$\mu$eV valley splitting orbital basis. For the quadrupole tensors we restrict ourselves to reporting the difference between the qubit states and the transition quadrupole for each case as this is the relevant part for $pO$ qubit operation.
\subsection{Dipole moments}
The full dipole moment vectors for all cases discussed in the main text are shown in Table.\ref{tab:dipole_moments}
\begin{table}[h]
    \centering
    \caption{Dipole moments of the qubit states $i=0$ and $i=1$, their
             difference $\Delta$, and the transition dipole $\mathbf{d}_{01}$
             (row ``01'') for FCI calculations at various valley splittings
             $E_\mathrm{VS}$, while the shell-filling picture is numerically
             calculated using the orbital states from the $240\,\mu$eV case.
             All values are in units of $e\cdot\mathrm{nm}$.}
   
    \begin{tabular}{l|l|S[table-format=-1.3]S[table-format=-1.3]S[table-format=1.3]|S[table-format=1.3]}
        \toprule
        Case & State
            & {$\mathbf{d}=(d_x,$}
            & {$d_y,$}
            & {$d_z)$}
            & {$|\mathbf{d}|$} \\
        \midrule\hline
        \multirow{4}{*}{$E_\mathrm{VS} = 2\,\mu\mathrm{eV}$}
            & 0        & -0.070 & -0.025 &  0.000 & 0.075 \\
        \cmidrule{2-6}
            & 1        & -0.070 & -0.026 &  0.000 & 0.074 \\
        \cmidrule{2-6}
            & $\Delta$ &  0.001 & -0.001 &  0.000 & 0.001 \\
        \cmidrule{2-6}
            & 01       &  0.000 &  0.000 &  0.000 & 0.000 \\
        \midrule\hline
        \multirow{4}{*}{$E_\mathrm{VS} = 14\,\mu\mathrm{eV}$}
            & 0        & -0.074 & -0.027 &  0.000 & 0.078 \\
        \cmidrule{2-6}
            & 1        & -0.074 & -0.027 &  0.000 & 0.078 \\
        \cmidrule{2-6}
            & $\Delta$ &  0.000 &  0.000 &  0.000 & 0.000 \\
        \cmidrule{2-6}
            & 01       &  0.000 &  0.000 &  0.000 & 0.000 \\
        \midrule\hline
        \multirow{4}{*}{$E_\mathrm{VS} = 38\,\mu\mathrm{eV}$}
            & 0        & -0.070 & -0.029 &  0.000 & 0.076 \\
        \cmidrule{2-6}
            & 1        & -0.070 & -0.027 &  0.000 & 0.075 \\
        \cmidrule{2-6}
            & $\Delta$ &  0.000 &  0.002 &  0.000 & 0.002 \\
        \cmidrule{2-6}
            & 01       &  0.000 &  0.000 &  0.000 & 0.000 \\
        \midrule\hline
        \multirow{4}{*}{$E_\mathrm{VS} = 50\,\mu\mathrm{eV}$}
            & 0        & -0.069 & -0.024 &  0.000 & 0.073 \\
        \cmidrule{2-6}
            & 1        & -0.078 & -0.023 &  0.000 & 0.081 \\
        \cmidrule{2-6}
            & $\Delta$ & -0.009 &  0.001 &  0.000 & 0.009 \\
        \cmidrule{2-6}
            & 01       &  0.001 & -0.007 &  0.000 & 0.007 \\
        \midrule\hline
        \multirow{4}{*}{$E_\mathrm{VS} = 60\,\mu\mathrm{eV}$}
            & 0        & -0.068 & -0.022 &  0.000 & 0.071 \\
        \cmidrule{2-6}
            & 1        & -0.077 & -0.023 &  0.000 & 0.080 \\
        \cmidrule{2-6}
            & $\Delta$ & -0.009 & -0.001 &  0.000 & 0.009 \\
        \cmidrule{2-6}
            & 01       & -0.003 &  0.007 &  0.000 & 0.007 \\
        \midrule\hline
        \multirow{4}{*}{$E_\mathrm{VS} = 66\,\mu\mathrm{eV}$}
            & 0        & -0.080 & -0.031 &  0.000 & 0.086 \\
        \cmidrule{2-6}
            & 1        & -0.074 & -0.025 &  0.000 & 0.078 \\
        \cmidrule{2-6}
            & $\Delta$ &  0.006 &  0.005 &  0.000 & 0.008 \\
        \cmidrule{2-6}
            & 01       & -0.003 &  0.005 &  0.000 & 0.006 \\
        \midrule\hline
        \multirow{4}{*}{$E_\mathrm{VS} = 85\,\mu\mathrm{eV}$}
            & 0        & -0.079 & -0.030 &  0.000 & 0.085 \\
        \cmidrule{2-6}
            & 1        & -0.077 & -0.023 &  0.000 & 0.081 \\
        \cmidrule{2-6}
            & $\Delta$ &  0.002 &  0.007 &  0.000 & 0.007 \\
        \cmidrule{2-6}
            & 01       &  0.001 & -0.002 &  0.000 & 0.002 \\
        \midrule\hline
        \multirow{4}{*}{$E_\mathrm{VS} = 210\,\mu\mathrm{eV}$}
            & 0        & -0.053 & -0.024 &  0.000 & 0.058 \\
        \cmidrule{2-6}
            & 1        & -0.051 & -0.012 &  0.000 & 0.053 \\
        \cmidrule{2-6}
            & $\Delta$ &  0.002 &  0.011 &  0.000 & 0.012 \\
        \cmidrule{2-6}
            & 01       & -0.005 &  0.001 &  0.000 & 0.005 \\
        \midrule\hline
        \multirow{4}{*}{$E_\mathrm{VS} = 240\,\mu\mathrm{eV}$}
            & 0        & -0.042 & -0.013 &  0.000 & 0.044 \\
        \cmidrule{2-6}
            & 1        & -0.048 & -0.018 &  0.000 & 0.051 \\
        \cmidrule{2-6}
            & $\Delta$ & -0.006 & -0.005 &  0.000 & 0.008 \\
        \cmidrule{2-6}
            & 01       & -0.002 &  0.004 &  0.000 & 0.005 \\
        \midrule\hline
        \multirow{4}{*}{Shell}
            & 0        &  0.000 &  0.000 &  0.000 & 0.000 \\
        \cmidrule{2-6}
            & 1        &  0.000 &  0.000 &  0.000 & 0.000 \\
        \cmidrule{2-6}
            & $\Delta$ &  0.000 &  0.000 &  0.000 & 0.000 \\
        \cmidrule{2-6}
            & 01       &  0.000 &  0.000 &  0.000 & 0.000 \\
        \bottomrule
    \end{tabular} \label{tab:dipole_moments}
\end{table}
\subsection{Quadrupole tensor}

This section contains the full quadrupole tensors differences between the two degenerate ground states, $\Delta Q = Q_{11} - Q_{00}$, are given for all cases with $E_\text{VS} \geq50\,\mu  \text{eV}$.
For $E_\text{VS} <50\,\mu  \text{eV}$, namely, $E_\text{VS} = 2$, $14$, and $38~\mu\text{eV}$, the quadrupole difference
vanishes to numerical precision (all components $< 0.1~e\cdot\text{nm}^2$) and is not listed.
The interacting electron case with $E_\text{VS} = 50~\mu\text{eV}$ gives
\begin{equation}
    \Delta Q^{(\text{FCI, 50})} =
    \begin{pmatrix}
         162 & -197 & 0 \\
        -197 & -162 & 0 \\
           0 &    0 & 0
    \end{pmatrix} e \cdot \text{nm}^2.
\end{equation}
The interacting electron case with $E_\text{VS} = 60~\mu\text{eV}$ gives
\begin{equation}
    \Delta Q^{(\text{FCI, 60})} =
    \begin{pmatrix}
          93 & -237 & 0 \\
        -237 &  -93 & 0 \\
           0 &    0 & 0
    \end{pmatrix} e \cdot \text{nm}^2.
\end{equation}
The interacting electron case with $E_\text{VS} = 66~\mu\text{eV}$ gives
\begin{equation}
    \Delta Q^{(\text{FCI, 66})} =
    \begin{pmatrix}
        -69 & 246 & 0 \\
        246 &  69 & 0 \\
          0 &   0 & 0
    \end{pmatrix} e \cdot \text{nm}^2.
\end{equation}
The interacting electron case with $E_\text{VS} = 85~\mu\text{eV}$ gives
\begin{equation}
    \Delta Q^{(\text{FCI, 85})} =
    \begin{pmatrix}
         45 & 251 & 0 \\
        251 & -45 & 0 \\
          0 &   0 & 0
    \end{pmatrix} e \cdot \text{nm}^2.
\end{equation}
The interacting electron case with $E_\text{VS} = 210~\mu\text{eV}$ gives
\begin{equation}
    \Delta Q^{(\text{FCI, 210})} =
    \begin{pmatrix}
          140 & 213 &   0 \\
        213 &  -140 &   0 \\
           0 &    0 & 0
    \end{pmatrix} e \cdot \text{nm}^2.
\end{equation}
The interacting electron case with $E_\text{VS} = 240~\mu\text{eV}$ gives
\begin{equation}
    \Delta Q^{(\text{FCI, 240})} =
    \begin{pmatrix}
          75 & -244 & 0 \\
        -244 &  -75 & 0 \\
           0 &    0 & 0
    \end{pmatrix} e \cdot \text{nm}^2.
\end{equation}
The shell-filling case using the orbital states gives
\begin{equation}
    \Delta Q^{(\text{SF})} =
    \begin{pmatrix}
        1187 &   199 & 0 \\
         199 & -1187 & 0 \\
           0 &     0 & 0
    \end{pmatrix} e \cdot \text{nm}^2.
\end{equation}

The quadrupole tensor elements of the transition densities are listed below with $E_\text{VS} = 2$, $14$, and $38~\mu\text{eV}$ having vanished to numerical precision (all components $< 0.1~e\cdot\text{nm}^2$) and are not included.
The interacting electron case with $E_\text{VS} = 50~\mu\text{eV}$ gives
\begin{equation}
    Q_{01}^{(\text{FCI, 50})} =
    \begin{pmatrix}
        -98 & -81 & 0 \\
        -81 &  98 & 0 \\
          0 &   0 & 0
    \end{pmatrix} e \cdot \text{nm}^2.
\end{equation}
The interacting electron case with $E_\text{VS} = 60~\mu\text{eV}$ gives
\begin{equation}
    Q_{01}^{(\text{FCI, 60})} =
    \begin{pmatrix}
        119 &   47 & 0 \\
         47 & -119 & 0 \\
          0 &    0 & 0
    \end{pmatrix} e \cdot \text{nm}^2.
\end{equation}
The interacting electron case with $E_\text{VS} = 66~\mu\text{eV}$ gives
\begin{equation}
    Q_{01}^{(\text{FCI, 66})} =
    \begin{pmatrix}
        123 &   34 & 0 \\
         34 & -123 & 0 \\
          0 &    0 & 0
    \end{pmatrix} e \cdot \text{nm}^2.
\end{equation}
The interacting electron case with $E_\text{VS} = 85~\mu\text{eV}$ gives
\begin{equation}
    Q_{01}^{(\text{FCI, 85})} =
    \begin{pmatrix}
        -125 &  23 & 0 \\
          23 & 126 & 0 \\
           0 &   0 & 0
    \end{pmatrix} e \cdot \text{nm}^2.
\end{equation}
The interacting electron case with $E_\text{VS} = 210~\mu\text{eV}$ gives
\begin{equation}
    Q_{01}^{(\text{FCI, 210})} =
    \begin{pmatrix}
         106 & -70 &  0 \\
        -70 & -106 &  0 \\
           0 &    0 & 0
    \end{pmatrix} e \cdot \text{nm}^2.
\end{equation}
The interacting electron case with $E_\text{VS} = 240~\mu\text{eV}$ gives
\begin{equation}
    Q_{01}^{(\text{FCI, 240})} =
    \begin{pmatrix}
        122 &   37 & 0 \\
         37 & -122 & 0 \\
          0 &    0 & 0
    \end{pmatrix} e \cdot \text{nm}^2.
\end{equation}
The shell-filling case using the orbital states gives
\begin{equation}
    Q_{01}^{(\text{SF})} =
    \begin{pmatrix}
        100 & -593 & 0 \\
       -593 & -100 & 0 \\
          0 &    0 & 0
    \end{pmatrix} e \cdot \text{nm}^2.
\end{equation}

\subsection{Moments between orbital basis} \label{app:BasisMoments}
\begin{figure}
    \centering
    \includegraphics[width=\linewidth]{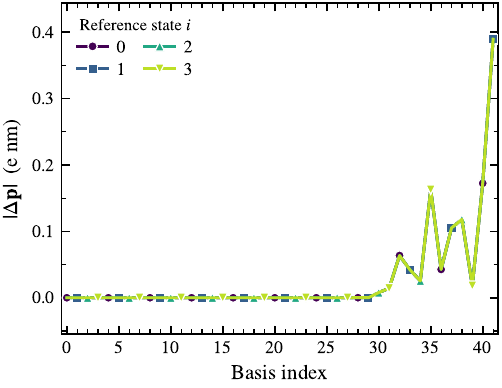} 
    \caption{Norms of the dipole moments between reference single-particle basis states, $i$, vs all single-particle states}
    \label{fig:DipoleBasis}
\end{figure}

\begin{figure}
    \centering
    \includegraphics[width=\linewidth]{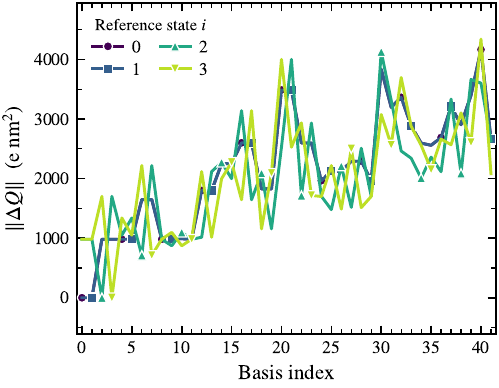} 
    \caption{Quadrupole tensor norm between reference single-particle basis states, $i$, vs all single-particle states}
    \label{fig:QuadrupoleBasis}
\end{figure}

The dipole and quadrupole tensor norm between the charge densities of the first four single-particle basis states and all other single-particle basis states are shown in Figs.~\ref{fig:DipoleBasis} and \ref{fig:QuadrupoleBasis}.

\section{Convergence}
A direct convergence test of the valley-resolved FCI calculations done by increasing our spatial basis size is beyond our computational resources. Preserving the rotational symmetry of the Hamiltonian requires adding complete shells when increasing state single-particle basis. Increasing the 42 state spatial-basis size used in the main text by one shell would mean adding the $i$ shell which is an additional 14 states (7 per valley). The Slater determinant basis size for five electrons grows as $N^5$ with the number of spatial orbitals and including 56 spatial orbitals exceeds our memory limits. 

In the high-valley-splitting regime, however, a reduced basis is possible. There the degenerate ground states relevant to the $pO$ qubit contain essentially no excited valleys because of the weak intervalley Coulomb coupling. They can therefore be removed from the basis, which corresponds to the limit $E_{\mathrm{VS}}\rightarrow\infty$, which closely approximates the high-valley-splitting regime. Removing the excited valley states halves the spatial basis size per shell, which allows us to extend the basis by two shells, up to the $j$ shell (36 spatial orbitals). These results are compared to the high-valley-splitting, $E_{\mathrm{VS}}=240\,\mu$eV, results of the main text. Overall, the results do not change significantly, and the two figures of merit are preserved; the dipole moments still vanish while the quadrupole tensor norm slightly increase. 

\begin{figure}[h]
    \centering
    \includegraphics[width=\linewidth]{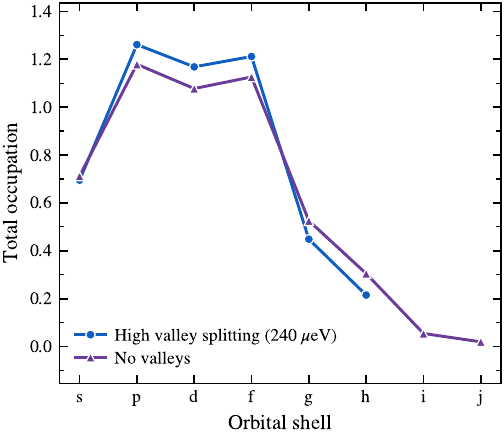}
    \caption{Orbital shell occupation of the multi-electron ground states for the FCI calculations with $E_\text{VS}=240\,\mu$eV ($s$ through $h$ shells, both valleys) and without valley states ($s$ through $j$ shells, lower valley only).}
    \label{fig:ShellOccupation_novalley}
\end{figure}

The shell occupation of the two calculations are shown in Fig.~\ref{fig:ShellOccupation_novalley}. The highest included shell, $j$, is occupied by only about 1\% of an electron, indicating that the charge density is close to converged and that including further shells would change it only minimally. 

\begin{figure}[h]
    \centering
    \includegraphics[width=\linewidth]{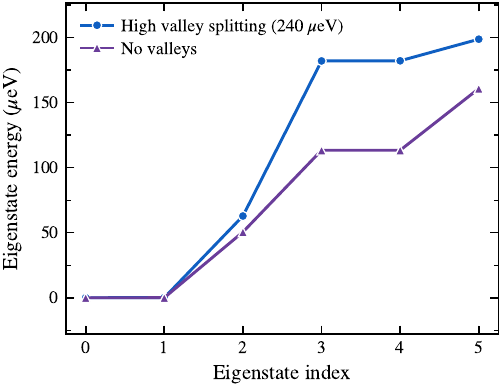}
    \caption{Energies of the six lowest energy eigenstates of Eq.~\eqref{eq:H5e} for the  calculations with $E_\text{VS}=240\,\mu$eV ($s$ through $h$ shells, both valleys) and without valley states ($s$ through $j$ shells, lower valley only).}
    \label{fig:FCIenergies_novalley}
\end{figure}

The eigenstate energies also only change minimally, and the degenerate ground-state doublet is preserved, as shown in Fig.~\ref{fig:FCIenergies_novalley}. The gap to the first excited state decreases from $63\,\mu$eV to $51\,\mu$eV.

The quadrupole tensors difference between the qubit states and the transition between them for the no-valley calculation are 
\begin{equation}
    \Delta Q^{(\text{FCI, no valley})} =
    \begin{pmatrix}
          71 & -298 & 0 \\
        -298 &  -71 & 0 \\
           0 &    0 & 0
    \end{pmatrix} e \cdot \text{nm}^2
\end{equation}
 and 
 \begin{equation}
    Q_{01}^{(\text{FCI, no valley})} =
    \begin{pmatrix}
        149 &   35 & 0 \\
         35 & -149 & 0 \\
          0 &    0 & 0
    \end{pmatrix} e \cdot \text{nm}^2.
\end{equation}
Their norms, $\|\Delta Q\| = 433\,e\cdot\text{nm}^2$ and $\|Q_{01}\| = 217\,e\cdot\text{nm}^2$, are both about $20\%$ larger than the high-valley-splitting regime results shown in Table~\ref{tab:NormQuadrupole}. The increase is consistent with larger quadrupole moments carried by the higher-shell basis states (Appendix~\ref{app:BasisMoments}). Since the Rabi frequency scales linearly with these norms, the estimate in Sec.~\ref{sec:Results} is a slight underestimate. 

The dipole moment difference $\Delta\mathbf{d}$ and the transition dipole moment $\mathbf{d}_{01}$ both remain below $0.008\,e\cdot\text{nm}$ which is within the numerical precision of the calculation. This is expected because removing the excited valleys preserves rotational and parity symmetries so the argument of Sec.~\ref{sec:SF_int_spec} still holds.

Finally, we note that removing the excited valleys by construction excludes the low-valley-splitting regime, whose ground doublet consists of valley excitations. The valley-resolved results in the main text remain essential to capture the two different regimes and the crossover point between them. 

\bibliography{references_2} 

\end{document}